\vskip .3in

 \centerline{\bf      Statistical Mechanics of Amplifying Apparatus}
\centerline {by Joseph F. Johnson}
\baselineskip=8pt 
\centerline {Math Dept., Univ.\ of Notre Dame}
\baselineskip=12pt
\vskip .2in

\centerline{\bf Abstract} \baselineskip=8pt 
\font\eightrm=cmr8
\eightrm

     We consider a toy model of a laser with negative temperature amplification.  
By passing to a new kind of thermodynamic limit, involving the renormalisation 
of Planck's constant to zero, we obtain a classically describable measurement  
apparatus.  We introduce a new definition of macroscopic observable which 
implements Bohr's insight that the observables of a measurement apparatus are 
classical in nature.  In particular, we obtain the usual non-abelian observables
 as limits of abelian, classical, observables.  This insight is made 
mathematically precise by carrying out Gibbs's program for Hamiltonian heat 
baths in our new context.  The method we use has a logical structure identical 
to that of Ford--Kac--Mazur in classical statistical mechanics.  Wigner's 
problem of quantum duality is then solved  by deriving the probabilities of 
measurement results from Schroedinger's equation.

\baselineskip=12pt 
\rm

\centerline{\bf Introduction}

      Wigner$^1$ wrote several fundamental analyses of the problem of quantum measurement, and 
the formulation of the problem which we adopt here is due to him, he called it the problem of 
quantum duality.  The deterministic axioms of 
non-Relativistic Quantum Mechanics, in Dirac's formulation, assume, to begin with, that every 
system is a closed system.  To every system is associated a Hilbert space ${\cal H}$, and a 
Hamiltonian operator $H$.  Every physical state of the system is described by a ray 
in that Hilbert space, as usual, we fudge the identifications and consider a non-zero wave 
function $\psi$ instead of the ray.  If the state of the system at time $t=0$ is $\psi_0$, 
then its state at time $t$ is given by 
          $$\psi_t=e^{{-itH}\over h} \cdot \psi_0 \in {\cal H}.$$

     But there are separate, probabilistic axioms for measurement processes.  For every 
observable, there is a self-adjoint operator $Q$ on ${\cal H}$.  If the system in the state 
$\psi$ undergoes a measurement process associated to this observable, the only possible 
results are the eigenvalues $\{\lambda_i\}$ of $Q$.  Assuming that $\psi$ is normalised 
and that its Fourier decomposition with respect to the normalised eigenvectors of $Q$ is 
           $\psi=\sum_i c_i v_i $, 
then the probability that the result will be $\lambda_i$ is $\vert c_i\vert^2$.  
If the result is $\lambda_i$, then the system is, as a result of the measurement process,
in the state given by the wave function $v_i$ even though such a transition or jump (called 
the reduction of the wave packet)  
violates Schroedinger's equation.

     On the other hand, the measurement apparatus itself must be a quantum system, possessed 
of a Hilbert space ${{\cal H}}_n$ to describe its physical states ($n$ is 
the number of particles in the apparatus) and a Hamiltonian $H_n$ to govern its time-evolution 
if it were in isolation.  Then the state space of the combined system of the microscopic, as 
we will call it, system originally under discussion and which is being measured, and the 
macroscopic, as we will call it, measuring apparatus, is ${\cal H}_n^{{\rm com}}={\cal H} \otimes {\cal H}_n$ 
and the joint Hamiltonian is the sum of the two Hamiltonians which would have governed each 
system in isolation plus an interaction term
(this is almost tautological)
         $$H_n^{{\rm com}}=H\otimes I_n + I\otimes H_n + H_n^{{\rm int}}.$$

The problem of quantum duality, Wigner$^1$, was that we have two rather different 
mathematical descriptions of the same physical process and it is not at all clear how to compare 
them.  In the former description, we are implicitly treating the measurement apparatus as if it 
were a classical system which did not obey the superposition principle so that we are sure that 
the result of the measurement process is always a definite pointer position, as it is called, 
a macroscopic pointer visibly and definitely pointing to one or another of the various possibilities 
$\lambda_i.$  Bohr (as quoted in J. Jauch, E. Wigner and M. Yanase$^2$) always insisted that the measurement apparatus had to be classically describable
and classical in nature.   Heisenberg (cf.\ J. Bell$^3$) always insisted that we have to put a cut somewhere, marking 
off when we use the first three axioms to analyse things, and when we use the measurement axioms.
 Von Neumann (as reported in Wigner$^1$) showed that as long as we do put the cut somewhere eventually, it makes no difference 
where we put it.  Wigner contributed to this analysis, emphasising that on the classical side of 
the cut will always be the observer's consciousness, at least.  So the methodology of introducing 
a cut used to be considered the solution to the problem of quantum duality.
But this no longer commands a consensus in light of advances in mesoscopic engineering, detection of quantum mesoscopic chaos, macroscopic superpositions of states, and so on.  

     Einstein asked the vague but profound question whether or not the probabilities of the latter 
three axioms did not arise from some underlying deterministic dynamics in an analogous way to 
the way they did in classical statistical mechanics.  He seems to have thought that this would 
mean either revising Schroedinger's equation or perhaps introducing hidden variables.  But the 
logically sophisticated treatment of classical statistical mechanics due to C. Darwin and R. Fowler$^4$, and 
extended by A. Khintchine$^5$, does not introduce hidden variables.  For us, the method by 
which G. Ford, M. Kac and P. Mazur$^6$ carry out the Gibbs program for the case of harmonic oscillators with a 
cyclic nearest neighbour interaction is paradigmatic.  (J. Lewis and H. Maassen$^7$ has extended this.)  By focussing Einstein's question on 
Wigner's formulation of the problem, we can answer it positively, taking Schroedinger's equation 
as the underlying deterministic dynamics.  That is, we derive the three probabilistic axioms 
from the three deterministic axioms.   We make no other assumptions$^8$ except that we have to 
introduce some sort of dictionary that compares quantum states with macroscopic pointer positions,
or else the comparison of the first three with the latter three axioms is logically impossible.
Basically, we push the cut out to infinity.  No real quantum system is exactly a measurement apparatus,
but the thermodynamic limit of quantum amplifying apparati becomes a classically describable 
measurement apparatus which exactly verifies the three probabilistic axioms in the limit as $n\rightarrow
\infty$.

     Physically, this model has much in common with aspects of previous work of 
H. Green$^9$ and A. Daneri, A. Loinger, and G. Prosperi$^{10}$.  
One important physical difference with Coleman---Hepp$^{11}$ is that the notion 
of macroscopic which we will introduce is the opposite of their notion of local observable
(which is from the theory of infinite volume thermodynamic limits, but not physically
appropriate here).  
The idea that coupling a Brownian mote to a negative 
temperature amplifier will amplify the motion of the 
mote from quantum motion, where observables do not 
commute, to classical motion, where the non-commutation 
becomes negligible, is perhaps due to J. Schwinger$^{12}$.  
Our model is that of a non-demolition measurement, 
however, and we assume that the amplifier exerts zero 
force on the incident particle.  

     This has mathematical similarities to the literature on the problem which takes the open system
approach.  But the physics is different.  We make the transition to irreversible classical stochastic 
dynamics a function of the coupling between the apparatus and the microscopic system.  
W. Zurek$^{12}$ and others$^{14}$ such as M. Collett, G. Milburn, and D. Walls$^{15}$ use a sort of open 
systems approach in that they put the interaction which 
is supposed to turn quantum amplitudes into classical probabilities in the coupling with 
the environment instead of in the coupling with the amplifier.  
We assume instead a closed joint system.  
 In principle, this difference should be detectable by experiment.
We also predict that the degree of validity of the probabilistic axioms should worsen as the size 
of the amplifying apparatus approaches the mesoscopic or even microscopic.  This should be 
detectable by experiment as well.  
The only real novelty is the notion of macroscopic which we introduce.  The need for a precise 
definition of macroscopic has long been felt.  Indeed, it is not possible to compare the two dual 
descriptions of the measurement process unless some sort of dictionary is provided.  Von Neumann 
and Wigner$^{16}$ (footnote 203) seems to have missed the need for a relatively sophisticated dictionary, and simply 
assumed that a macroscopic pointer position 
corresponded, more or less approximately in the strong topology on ${\cal H}_n$, with a large set 
of wave functions.  It seems to be this unquestioned assumption which has been the obstacle to 
progress along Wigner's original lines.  Abandoning it may mean the abandonment of the ancient dream 
of psycho-physical parallelism.$^{16}$  

     In connection with this I quote P. Dirac$^{17}$.  ``And, I think it might turn out that ultimately Einstein will prove to be right, \dots that it is quite likely that at some future time we may get an improved quantum mechanics in which there will be a return to determinism and which will, therefore, justify the Einstein point of view.  
But such a return to deteminism could only be made at the expense of giving up some other basic idea which we now asume without question.  We would have to pay for it in some way which we cannot yet [1977] guess at, if we are to re-introduce determinism.''  
In this regard I wish to point out that there is almost no credible evidence in favour of the hoary 
psycho-physical parallelism, and there is a good deal of evidence for the time-dependent form of 
Schroedinger's equation's being absolutely linear and unitary.  

     It will be noticed that our physical interpretation, then, is that there are only waves.  There are 
no particles.  The problem of quantum duality, then, becomes the problem of calculating, relatively 
explicitly, how it is that the interaction of the wave of, say, a microscopic system with one degree 
of freedom (called, sentimentally, an incident particle) interacts with the wave of the amplifiying 
apparatus to sometimes (with a definite probability) produce a ``particle-event,'' i.e., the registering 
of a loud click on the part of tha apparatus, and other times, produce no such detection event.  Thus the 
seeming particle-events, the seeming detection of particles, is an artifact of the amplification process 
in the interaction of two waves.  
     
     Feynman, famously, did not think so, but he did think that perhaps a little more could be said 
about the problem of wave-particle duality,  
``We and our measuring instruments are part of nature and so are, in principle, described by an amplitude function satisfying a deterministic equation.  
Why can we only predict the probability that a given experiment will lead to a definite result?
From what does the uncertainty arise?  Almost without a doubt it arises from the need to amplify the effects of single atomic events to such a level that they may be readily observed by large systems.''  

``\dots In what way is only the probability of a future event accessible to us, whereas the certainty of a past event can often apparently be asserted?\dots Obviously, we are again involved in the consequences of the large size of ouselves and of our measuring equipment.  
The usual separation of observer and observed which is now needed in analyzing measurements in quantum mechanics should not really be necessary, or at least should be even more thoroughly analyzed.  
What seems to be needed is the statistical mechanics of amplifying apparatus.''$^{18}$  

It is not necessary to think about the philosophical meaning of probability, as A. Sudbery$^{19}$ 
remarks, 
``It is not in fact possible to give a full definition of probability in elementary physical terms.''
``Attempts to define probability more explicitly than this are usually either circular \dots or mysterious \dots This is not to say that the question of what probability means, or ought to mean, is not interesting and important; but the answer to that question, if there is one, will not affect the properties of probability that are set out here, and we can proceed without examining the concept any further.''
All that is necessary is to straightforwardly imitate the classical methods of statistical 
mechanics, in a physically appropriate way.  Remarkably, the result we arrive at corresponds exactly to J. von Plato's philosophically motivated amendment$^{20}$ of the traditional frequency theory of probability.  The naive frequency theory suffered from grave logical circularities and von Plato fixed these by incorporating the procedures of classical statistical 
mechanics.  His work has not received the attention it deserves because he had to rely on 
the underlying dynamics' being deterministic, so it was usually assumed that his purely 
logical, almost antiquarian concerns, could not be relevant to a quantum world.  This turns out to be a misapprehension.  

\centerline{\bf The Concrete Model}

     The phenomena of interest can be exhibited with finite-dimensional Hilbert spaces.  
Let the state space of an incident particle be ${\bf C}^2$.  We consider that this space has
basis $\psi_0$, which means the particle is somewhere else, and $\psi_1$, which 
means the particle is present.  For each $n$,
${\cal H}_n$ is the Hilbert space of wave functions describing the state space of an 
$n$-particle system which is an amplifying device.  Since it is a quantum system, it is not 
yet exactly a measuring apparatus.  It does not become such until the passage to the limit. 
We let ${\cal H}_n$ be the $n$-fold tensor product of two-dimensional Hilbert spaces with 
bases $|1\rangle $ and $|0\rangle $.  We write $|1\rangle \otimes|1\rangle \otimes|0\rangle = |110\rangle $, for 
example.  Think of this as the 
state where the first two particles on the left of the system are in an excited state, and 
the right-most particle is in a discharged state.

    In the presence of an incident particle in the state $\psi_1$, the amplifying apparatus 
will evolve in time under the influence of a cyclic nearest-neighbour interaction which is 
meant to model the idea of stimulated emission or a domino effect.  In the absence of a 
particle, the dynamics on the amplifying device will be trivial.  We can afford 
to be so schematic because the behaviour of the thermodynamic 
limit is extremely robust with respect to inessential variations in the definitions of the 
dynamics and interactions.  For each $n$, we define the Hamiltonian operator $H_n$ which 
governs the cyclic nearest neighbour interaction on the amplifying device as follows.  

The picture is of $n$ particles in a line.  Each one passes 
on its state to its rightmost neighbour, and the first one receives its state from the one
on the right end (this is the cyclicity).  This takes place in time $ {1\over n}$.  If 
$\vert \pi_1 \pi_2 \dots \pi_n\rangle $ is a typical basis vector, we have that it transforms into 
$\vert \pi_n \pi_1 \dots \pi_{n-1}\rangle $ after $ {1\over n}$ units of time.   Assume for simplicity
that $n$ is prime.  Then there are, besides $|000\dots 0\rangle $ and $|111\dots 1\rangle $, 
${{2^n-2}\over n}$ (which is an integer by Fermat's little 
theorem)   
invariant $n$-dimensional subspaces of ${\cal H}_n$ for example, one is spanned by 
$|100\dots 0\rangle $, $|010\dots 0\rangle $, and so on.  
We give the restriction of ${\cal H}_n$ to any one of these subspaces (they are all
isomorphic).  It is a Vandermonde matrix with eigenvalues $0,h,2h,3h,\dots,(n-1)h$.  

The intuitive picture is that this device is getting more and more classical as $n$ goes 
to infinity.  So the energy levels get closer and closer, approaching a continuum, the 
particles get closer and closer which is why the interaction, at a constant speed, 
travels from a particle to its neighbour in less and less time, the mass of each particle 
decreases as $ {1\over n}$, so that we have finite length and fixed density and fixed total 
energy.  We must, therefore, renormalise, letting $h \propto {1\over n}$.

We must couple the amplifier to the incident particle.  
The Hilbert space of the combined system is ${\cal H}_n^{{\rm com}}={\bf C}^2\otimes{\cal H}_n
=<\psi_0> \otimes {\cal H}_n \oplus <\psi_1> \otimes 
{\cal H}_n$.  So we need only define $H_n^{{\rm com}}$, the Hamiltonian of the joint system, by 
giving it on the first factor, where it is trivial, and on the second factor, where it is
$H_n$.  This is the explicit toy model of a quantum amplifier, we have now to study the 
question, what quantities at each finite stage correspond, in the limit, to a macroscopic 
pointer position of a classical measurement apparatus?

We need a precise notion of what is a macroscopic system, in terms of the axioms of 
quantum mechanics, and what is a macroscopic observable. 
Although our model owes a great deal to the Coleman--Hepp model, this notion is the exact 
opposite of Hepp's notion of a local observable (which is also the usual one in the infinite
volume thermodynamic limit of Haag, Ruelle, and others$^{21}$).  We wish to implement the intuition 
of a function on the phase space which cannot distinguish between states which differ from 
each other in a finite or negligible number of spots.  

In the classical methods of thermodynamics, one worked essentially with one 
observable at a time and there was really a sequence of them for 
each $n$.  I.e., for each $n$, one has $f_n$ a physically significant phase function on 
the space ${\cal H}_n^{{\rm com}}$, or its classical analogue.  And the physical significance is the 
same as $n$ varies.  It could be total energy of a part of the system, for example the 
Brownian mote.  For us, it could be a formalisation of some intuitive idea such as, the 
percentage of excited particles in the left half of the device.  
These methods are acutely, if disparagingly, described by R. Minlos$^{21}$.  
``For a long time the thermodynamic limit was understood and used too formally: the mean values of some 
local variables and some relations between them used to be calculated in a finite ensemble and then, in the formulas obtained, the limit passage was carried out.''
Excellent agreement with experimental results were obtained that way.
At any rate, we formally define such a sequence of $f_n$ to be macroscopic if whenever the 
sequence of norm one vectors $v_i\in{\bf C}^2$, $i>0$, satisfies 
$$\lim_{n\rightarrow\infty} f_{n+n_o}(v_o\otimes v_{n_o+1}\otimes v_{n_o+2}\otimes\dots\otimes v_{n_o+n})$$
exists for some $n_o$ and some $v_o\in{\cal H}^{{\rm com}}_{n_o}$, then it is independent of the choice of
$n_o$ and $v_o$.  

We now define the family $f_n$, which in the limit, becomes the pointer position of the 
measuring apparatus.  There is a basis of ${\cal H}_n^{{\rm com}}$ consisting of separable vectors
 of the form $\psi_{\pi_0}\otimes\vert \pi_1 \pi_2 \dots\pi_n\rangle $ where, as before, the 
$\pi_i$ are 0 or 1.  Let $C$ be the set of basis vectors such that all but a negligible 
number of the $\pi_i$ for $i<n/2$ are 1 and all but a negligible number of the others are 0.
(This is the device being `cocked' and ready to detect.  It is very far from being a 
stable state, in the limit.)  By negligible, we mean that as a proportion of $n$, it 
goes to zero as $n$ increases.  For $v_n$ any state of the combined system, and we may 
take $v_n$ normalised of length unity,  let $c_i$ be the Fourier coefficients of $v_n$ with 
respect to the cocked basis vectors, i.e., those in $C$.  Define $f_n(v_n)=1-\sum_i|c_i|^2$.

\centerline {\bf Passing to the Thermodynamic Limit: the Ming Effect}

Our procedure is modelled on that of Ford--Kac--Mazur, which indeed is modelled on the 
usual understanding of the Gibbs program.  However, a major physical difference is the 
renormalisation which we introduce.  Another is that the physical quantity whose 
limit they study is an autocorrelation; ours is a macroscopic pointer position.
Now we are interested in phenomena in the limit as $n$ approaches $\infty$, yet one cannot 
directly compare a value of $f_n$ on a vector $v_n$ with $f_{n+1}(v_n)$.  
In keeping with the procedures of classical statistical mechanics, 
one compares time or phase averages of the various $f_n$ as $n$ varies.  
(Phase averages would be taken over the submanifold of accessible states, it is more convenient
for us to deal with time averages.  Time averages have been made the basis for von Plato's theory 
of the meaning of probability statements.)
Let the incident particle
be in the state described by any (normalised) wave function in ${\bf C}^2$.  Let it be 
$v_0=a_0\psi_0+a_1\psi_1$.  The amplifier is in the state $|111\dots 000\rangle  $ in $C$.  
We now calculate the 
limit, as $n$ approaches $\infty$, of $<f_n>$, 
where $<f_n>$ means the time average of $f_n$ taken over a typical trajectory in the manifold 
of accessible states inside of ${\cal H}_n^{{\rm com}}$.

We will then find a classical dynamical system $\Omega_\infty$ which has a mixed 
state $X$, which depends on $v_0$, and a classical dynamical variable $F$ whose 
expectation values match these limits.  At any rate, it is elementary to calculate 
$\lim_{n\rightarrow\infty} <f_n>$, it is $|a_1|^2$.
(The microscopic indident particle triggers a domino effect or macroscopic `flash' which is
`bright and clear' i.e., `Ming'.)

The procedure of Ford--Kac--Mazur was to consider a dynamical variable 
$f_n$ on each finite system (classical), $M_n^{{\rm com}}$.  It was the 
position of the Brownian mote. 
They then took a kind of phase average of this, its autocorrelation function, 
$g_{t,n}$.  (In their original paper, they placed a carefully chosen probability distribution on the heat bath: it has since been shown that their results are insensitive to the 
choice of the mixed state of the heat bath.)  They then passed to the limit as $n$ approaches $\infty$ 
using a suitable renormalisation.  They then searched for a dynamical 
system and a stationary Gaussian stochastic process on that system which 
would have as its autocorrelation function this limit.  
(Furthermore, K. Hannabuss's use$^{22}$ of the Sz.\ Nagy dilation is in much the same tradition: the dilation is 
constructed simply from the correlation functions.)
We will carry 
this out in our setting (but using time averages for convenience).

We search for $\Omega_\infty$, $F$, and $X_{v_0}$ as above, satisfying 
$$\int_{\Omega_\infty}FdX_{v_0}=\lim_{n\rightarrow\infty}<f_n>.$$
Let the state of the (classical limit) measurement apparatus where the pointer position 
points to cocked (and hence, an absence of detection) be the point $P_0$.  Let the state 
where the excited states of the apparatus are proceeding from out of its initial cocked 
state, and travelling steadily towards the right, be $P_1$.
Then $\Omega_{\infty}=\{P_0,P_1\}$.  The dynamical variables on this space are generated by
the characteristic functions of the two points, $\chi_{P_0}, \chi_{P_1}$. 
Let $F$ be $\chi_{P_1}$. It is the pointer position which registers detection.
The mixed state of  $\Omega_{\infty}=\{P_0,P_1\}$ which gives the right answer   
when the incident particle is in state $v_0$ is the probability distribution which gives 
$P_0$ the weight $|a_0|^2$, and $P_1$ the weight $|a_1|^2$.  
This is precisely what it means to say the the measuring apparatus will register the 
presence of the particle with probability $|a_1|^2$, and its absence with probability 
$|a_0|^2$.  

\centerline{\bf Summary}

     This answers a famous question of Einstein's partly
 affirmatively and partly negatively.  It is possible to 
derive the probabilities in measurement results from an 
underlying deterministic dynamics analogously to the way
 it was done in classical statistical mechanics of Einstein's day.  It is not necessary to assume that quantum
 mechanics is incomplete in order to do this: we may take
 the wave function as the complete description of nature 
and Schroedinger's equation as exactly and universally 
valid.

This analysis shows that wave-packet reduction in the strong topology, 
even approximately, need not occur.  
It also shows that the transmutation of quantum amplitudes into classical 
probabilities depends only on the macroscopic nature of the pointer position  
and its coupling to the microscopic system being measured.  It does not depend 
on any back-force being exerted on the microscopic system.  
The coupling can be as gentle, in its effect on the microscopic system, as desired.
It suggests that the degree of 
validity of `measurement,' as an approximation to a physical amplification 
process, depends on the size of the apparatus.  Mesoscopic 
amplifiers should, then, demonstrate detectable noise phenomena 
in comparison to macroscopic amplifiers.

     This analysis further shows that it is not necessary to invoke the effect 
of the environment in order to construct a logically coherent theory of 
decoherence.  The fact that probabilities arise even from an amplifier which is
 in a pure state shows that quantum measurement can be explained without 
super-selection rules.  Thus the question whether the observed behaviour of 
measurement processes is due to the coupling between the apparatus and the 
microscopic system, or due to the coupling between the apparatus and the 
environment$^{23}$, becomes a question for experiment.    

\centerline{\bf References}

\noindent [1] E. Wigner, Z. Phys.\ {\bf133} (1952), 101; Am.\ J. Phys.\ {\bf31} (1963), 6.

\noindent [2] J. Jauch, E. Wigner and M. Yanase, Nuovo Cimento {\bf48} (1967), 144.

\noindent [3] J. Bell, Helv.\ Phys.\ Acta {\bf48} (1975) 447; Physics World {\bf3} (1980), 33.

\noindent [4] C. Darwin and R. Fowler, Philos.\ Mag.\ {\bf44} (1922), 450; 823; Proc.\ Cambridge Philos.\ Soc.\ {\bf21} (1922), 391.

\noindent [5] A. Khintchine, {\it Mathematical Foundations of Statistical Mechanics}, New York, 1949.

\noindent [6] G. Ford, M. Kac and P. Mazur, J. Math.\ Phys.\ {\bf6} (1965), 504.

\noindent [7] J. Lewis and H. Maassen, {\it Lecture Notes in Mathematics} {\bf1055}, Berlin, 1984, pp.\ 245-276.

\noindent [8] E. Farhi, J. Goldstone and S. Gutmann, Ann.\ Phys.\ (NY) {\bf192} (1989), 368.

\noindent [9] H. Green, Nuovo Cimento {\bf9} (1958), 880.  

\noindent [10] A. Daneri, A. Loinger and G. Prosperi, Nucl.\ Phys.\ {\bf33} (1962), 297.   

\noindent [11] K. Hepp, Helv.\ Phys.\  Acta {\bf45} (1972), 237.

\noindent [12] J. Schwinger, J. Math.\ Phys.\  {\bf2} (1961), 407.

\noindent [13] W. Zurek, Phys.\  Rev.\  D {\bf24} (1981), 1516; {\bf26} (1982), 1862; Physica Scripta {\bf76} (1998), 186.

\noindent [14] C. Gardiner and P. Zoller, {\it Quantum Noise}, Berlin, 2000, pp.\ 212-229.

\noindent [15] M. Collet, G. Milburn and D. Walls, Phys.\ Rev.\ D {\bf32} (1985), 3208. 

\noindent [16] J. von Neumann, {\it Mathematicsche Grundlagen der Quantenmechanik}, Berlin, 1932, p.\ 223.

\noindent [17] P. Dirac, {\it Directions in Physics}, New York, 1978, p.\ 10.

\noindent [18] R. Feynman and A. Hibbs, {\it Quantum Mechanics and Path Integrals}, New York, 1965, p.\ 22.

\noindent [19] A. Sudbery, {\it Quantum Mechanics and the Particles of Nature}, Cambridge, 1986, pp.\ 41ff.

\noindent [20] J. von Plato, ``Ergodic Theory and the Foundations of Probability,'' in {\it Causation, Chance, and Credence, Proceedings of the Irvine Conference on Probability and Causation}, edited by B. Skyrms and W. Harper, vol.\ 1, Kluwer, 1988, pp.\ 257-277.

\noindent [21] R. Minlos, {\it Introduction to Mathematical Statistical Physics}, Providence, 2000, p.\ 22.

\noindent [22] K. Hannabuss, Helv.\ Phys.\ Acta {\bf57} (1984), 610; Ann.\ Phys.\ (NY) {\bf239} (1995), 296.

\noindent [23] H. Zeh, in {\it Proceedings of the II International Wigner Symposium}, Goslar, Germany, 1991, edited by H. Doebner, W. Scherer, and F. Schroeck, Jr., World Scientific, 1993, 205.

\noindent [24] J. Johnson, in {\it Quantum Theory and Symmetries, Proceedings of the third International Symposium,} Cincinnati, 2003, edited by P. Argyres \it et al\rm., Singapore, 2004, 133.
\end